\documentclass{article}
\begin{document}

\begin{center}
\centerline{\large \bf Physical origin of nonlinear phenomena in optics.}
\end{center}

\vspace{3 pt}
\centerline{\sl V.A.Kuz'menko}
\vspace{5 pt}
\centerline{\small \it Troitsk Institute for Innovation and Fusion 
Research,}
\centerline{\small \it Troitsk, Moscow region, 142190, Russian 
Federation.}
\vspace{5 pt}
\begin{abstract}

	 The physical nature of numerous of the nonlinear phenomena 
in optics is explained by inequality of forward and reversed optical 
transitions, that corresponds to a principle of time invariance 
violation in electromagnetic interactions. The direct and indirect 
experimental proofs of such inequality and direction of the further 
researches are discussed.

\vspace{5 pt}
{PACS number: 42.50}
\end{abstract}

\vspace{12 pt}
\section{Introduction}

	 The research of the physical phenomenon usually pursues two 
purposes. The first purpose is to explain qualitatively a nature of 
the phenomenon and the second one is to describe it quantitatively 
with the help of suitable mathematical model. These tasks frequently 
have a different degree of complexity and its not always can be 
executed simultaneously. It happens, that the explanation of physical 
effect is rather simple, and its quantitative description is very 
difficult. It is a case with a prediction of weather. 

	 There are cases when the phenomenon appears easier to describe 
quantitatively, than to understand its physical nature. The periodic 
law is such vivid example. It is the main law of chemistry. The 
periodic table was created by Mendeleyev on the base of atomic 
weights. It gives excellent description of elements and their compounds 
properties. But this law for very long period had not a physical
explanation. Only there are a lot of years later, after discovery 
of electrons, protons, neutrons and creation of the quantum mechanics,
the physical essence of the law has become clear.

	 The situation, similar to this case, probably, has a place 
now in optics. Many years the mathematical description of 
dynamics of optical transitions exists. It is based on the Bloch equations,
which was offered in 1946 for the description of a nuclear magnetic 
resonance [1]. Now Maxwell-Bloch equations are used for the 
description practically all nonlinear effects in optics. Such model 
gives really good description of the optical phenomena [2, 3]. 
However the Bloch equations have not clear physical interpretation. 
Therefore there are large difficulties at attempts to understand 
physical sense of such descriptions. For this purpose such concepts,
as coherence, interference and a superposition of states are 
usually used. But its nothing explain. Its look like as attempt of 
physical interpretation of the mathematical description, when to 
operations of addition and subtraction of the members in the 
equations there correspond principles of interference and states 
mixing.

	 In present article the physical explanation of set of the 
phenomena in optics is discussed. It is based on a principle of 
inequality of forward and reversed optical transitions. However, 
for a long time and strongly settled opinion exists that such optical 
transitions are equivalent, and electromagnetic interactions as a
whole in nature are time invariant [4]. It is difficult to 
understand what is the base of this opinion. Usually it is 
referred to Einstein's opinion "that physics could be 
restricted to the time-symmetric case for which retarded and 
advanced fields are equivalent" [5]. 

	 The basis of this opinion can not be the equality of Einstein 
coefficients for absorption and stimulated emission of photons. 
Einstein coefficients characterize the integrated cross-section 
of optical transition. The preservation of time invariance demands 
not only equality of integrated cross-sections, but also equality 
of spectral width of direct and reversed optical transitions. Thus, 
the Einstein coefficients have not the direct relation to invariance
of process of photon absorption. 

	 The experiments on study of differential cross-section of 
reaction $Al^{27} + p = \alpha + Mg^{24}$ also can not be considered as 
the proof of time invariance preservation [6]. These experiments 
show equality of differential cross-sections of forward and backward 
processes. But we are interested more in comparison of 
cross-sections forward and reversed processes. It is a case, when 
the backward process follows directly after the forward one and 
the system can have memory about the forward process. So, the 
experimental proofs of time invariance preservation in 
electromagnetic interactions are absent. 

	 From other side the opinion exists, that (except for the 
case of $K^{0}$-meson decay) there are absent also the experimental 
proofs of time invariance violation [7]. This is error. In optics 
there is a set of the experimental proofs of time invariance 
violation in electromagnetic interactions. But usually there are 
the indirect experimental proofs, which essence is not realized 
to the present time.

	 The problem is that we are not able to measure separately 
the parameters of forward and reversed processes. Under action of 
radiation both forward and reversed processes proceed simultaneously.
A width of optical transition is usually connected with lifetime of 
the exited state. Therefore, when wide optical transitions are used,
some temporary difficulties appear. If the lifetime of exited state 
is big, the line of optical transition is very narrow and 
difficulties with inhomogeneity of a spectrum exist. As a 
result the information on parameters of the reversed 
optical process has mainly indirect character. For example, the 
Autler-Townes effect [8] and the numerous examples of amplification 
without inversion [9] are the indirect proofs of inequality of 
forward and reversed processes. 

	 However the brightest indirect proof of this inequality in 
our opinion is the effect of adiabatic population transfer in 
two-level system due to sweeping of resonant conditions [10]. 
When the resonance radiation interacts with the two-level system, 
the so-called periodical Rabi oscillations of the level population 
takes place. But if the sweeping of resonance conditions appears 
(for example, the frequency of radiation is changed), the full 
population transfer from the initial level to the opposite one 
takes place. And this result does not depend on intensity of 
radiation (if it is rather strong). This surprising result is 
well described in Bloch model. It is said, that the physical 
nature of this effect can hardly be explained verbally in simple
terms, but one should carefully follow the behavior of the vectors 
in the model of the rotating wave [11]. This situation resembles 
very much an appeal of the prestidigitator asking the spectators 
to watch his hands carefully while he is making the manipulations.
And the result seems to be the same. An explanation is given, 
but its physical essence is absolutely unclear. 

	 Here it is important that the physical explanation of this 
effect is impossible, if we assume equality forward and reversed 
processes. Its should be something different, that the atom could 
know, what level is initial, and what level is final. It is 
natural to expect, that the difference can consist in different 
width and cross-section of forward and reversed transitions. The 
rapid adiabatic passage effect is the most convincing indirect 
proof of inequality forward and reversed processes.

	 However for the decision of a problem only indirect proofs 
are insufficient. The direct proofs are necessary. And such direct
proof exists. It is connected to physical object, which has an 
unusual combination of properties: extremely large homogeneous width 
of optical transition is combined with the big lifetime of the 
exited state toward to spontaneous emission. In this case it appears 
very easily experimentally to find out the large difference between 
parameters of forward and reversed processes. This unusual object 
is the so-called wide component of line in absorption spectrum of 
polyatomic molecules. The following section is devoted to 
discussion of experiments with this object. 

\section{ Infrared laser multiple photon excitation of polyatomic
 molecules.}

	 The phenomenon of the infrared multiple--photon excitation 
(IR MPE) and collisionless dissociation of polyatomic molecules 
was discovered in works [12, 13]. It was founded, that polyatomic
molecules can absorb tens photons of laser radiation and dissociate 
without collisions. Numerous works were carried out later aimed to 
clarify the mechanism of this process.  This interest was 
stimulated by the fact, that the widths both of laser radiation 
and molecule absorption lines are substantially lower, than the 
anharmonicity of molecular vibrations. It means, that the absorption 
of second quantum of laser radiation should not occur. 
 
	 For an explanation of the mechanism of process in the 
former work the hypothesis about existence of so-called 
"quasicontinuum" of vibrational states was proposed. Despite 
of the argued criticism [14] such idea has received the broadest 
distribution. In the recent years, however, the views on the nature 
of "quasicontinuum" have changed dramatically. Earlier, it was 
accepted, that "quasicontinuum" consists of a huge number of narrow 
lines arising as a result of coupling different vibrational states. 

Now it is widely believed [15], that the absorption line is unique, 
but it becomes very wide. The origin of the "quasicontinuum" now 
is bounded up with intramolecular vibrational relaxation (IVR) 
process. This is a reasonable idea. The IVR process can be very fast
(picoseconds timescale). The corresponding Lorentzian width of the 
absorption line can be in this case comparable with anharmonicity 
of the molecular vibrations. The main disadvantage of this model 
is that it does not explain how the molecules can be excited in 
the region of low vibrational levels, where the IVR is absent and 
the absorption lines remain narrow. Experiments show, that 
excitation of molecules in this area occurs without essential 
difficulties, but the theory gives no satisfactory explanation 
of this fact.

	In works [16, 17] an idea was proposed, that the IR MPE 
process is a trivial result of absorption in the area of line 
wings, but antiviral is the nature of these wings. Practically, 
the possible role of line wings was not discussed in the 
literature earlier. It is, apparently, due to the fact, that 
appropriate estimations can easily be made.  The lifetime of 
excited states of molecules due to spontaneous emission in the 
infrared region lays in the millisecond timescale.  The natural 
width of line must to be smaller than $100 Hz$. Even for the 
strongest molecular transitions, at the distance from the line 
center equal to the value of molecular anharmonicity, the 
Lorentzian contour of the natural width would have an absorption 
cross--section smaller then $\ 10^{-25}\ cm^2$. This cross--section 
cannot play any appreciable role in overcoming the anharmonicity 
of molecular vibrations.

	However, such estimation has not been tested in experiment 
earlier. It is possible to assume, that for some unknown reasons, 
intensity of real line wings is much higher, than the theory predicts. 
How high the intensity of line wings should be to explain 
the observable effect of laser excitation of molecules? 
Rather correctly such information can be derived from 
the experimental results of works [18-20], where the depletion of 
rotational states of  $SF_6$ molecules by TEA $CO_2$--laser 
radiation was studied in the conditions of molecular jet. 
The results of such processing are presented in Fig.1.
Except for the usual narrow component of the line with a Doppler
width $\sim 30 MHz$, the wings, or more precisely speaking, 
a wide component of the line should exist with a
cross--section $\sigma\simeq 6\cdot{10^{-20}} \ cm^2$ and with 
a Lorentzian full width at half medium $\sim 4.5\ cm^{-1}$. 
The relative integral intensity of this component is rather small, 
$\sim 0.2\% $, but it is high enough for efficient excitation of 
molecules from all rotational states.

	 For experimental test of existence the wide components of a 
line the form of line should be studied on the large depth. This 
strongly prevents by inhomogeneous broadening, which connected 
with distribution of molecules on different rotational states. 
At room temperature the dense spectrum of transitions from 
different rotational states is observed. But here it is 
important to pay attention that Lorentzian contour is 
rather flat and wide component of lines can manifest itself as 
far natural wings of absorption bands. There are many 
publications about study of far wings of absorption bands 
of small and light molecules [21,22]. These wings are the result 
of collisional broadening of absorption lines. For heavy 
polyatomic molecules in a gas phase the far wings of absorption 
bands of other nature were discovered [23]. The experiments have 
shown, that the cross-section of absorption in the region of these 
wings does not depend on pressure of gas. So, its have a natural
nature.

In Fig.2, the spectral dependence of the absorption cross-section 
of $SiF_4$ molecules around the $\nu_3$ absorption bands is presented. 
The edges of the absorption band have approximately an exponential 
form, the slope being greater for the blue side, than for the red one. 
At the distance more, than 25---40 $cm^{-1}$ from the band center, 
much more flat wings are observed. The curve (2) is a Lorentzian 
profile with FWHM $=4.5 cm^{-1}$, which passes through the 
point with minimal absorption cross-section in the given spectral 
range. So, we can see, that the far band wings have a Lorentzian behavior.

In Fig.3 a spectral dependence of the absorption cross-section of 
$SF_6$ molecules is shown. In this rather typical case the far band 
wings are masked by intense combination bands.

	 The measurement of intensity of far absorption band wings 
allows to estimate integrated intensity of wide components of lines. 
The same value can also be experimentally estimated by other 
method: on the data about saturation of absorption spectrum of 
polyatomic molecules by radiation of pulse $CO_2$- laser at low gas 
pressure  [23]. The experiments show, that the relative intensity 
of wide component quickly grows with increasing of number of atoms 
in a molecule and branching degree of the molecules. Thus the estimated 
average relative integral intensity of the line wings at room 
temperature varied from $\sim 0.6\%$  for  $SF_6$ and $SiF_4$  
to $\sim 90\%$ for $(CF_3)_2O$ and  $(CF_3)_2CO$. 

	 The nature of wide components of lines is unknown. As a 
hypothesis the following explanation is offered. A certain 
mechanism of averaging of the rotational moment of molecule 
inertia works during the vibrational motion of atoms. In the 
large molecules this mechanism undergoes periodic and convertible 
breaking. As a result the absorption line splits on a clump of 
narrow lines, and the short-lived moments of breaking correspond 
to a wide component of a line [24]. 

	 The wide component of line is unique physical object, which 
has the long lifetime of the excited states toward to spontaneous 
emission and large homogeneous spectral width of optical transition. 
This combination of properties is very convenient for study the 
reversed optical transition in conditions of a molecular beam.

	 A wide component of lines was easily observed in work [25] 
at study the absorption of radiation of continuous $CO_2$- laser 
in a molecular beam with cryogenic bolometer. Rotational 
temperature in a molecular beam is very low ( $\sim 5^{0}K$ ). 
It radically changes character of a molecule absorption spectrum. 
The absorption lines become very rare and the $CO_2$- laser
radiation interacts practically only with the wide component 
of lines. Unfortunately, the authors had not understood with what 
thing they deal with and later the work with this object was closed. 

	 In present case we are not interested in a line wings itself, 
but in the results of double optical resonance experiments in a molecular 
beam [25]. For the first laser beam the absorption spectrum represents 
wide continuum. For the second laser beam, which cross the 
molecular beam later, besides this wide continuum a sharp dip 
with a width   $\sim 450 kHz$ is observed. It characterizes a 
spectrum of the reversed optical transition. The ratio of forward 
and reversed optical transitions widths exceeds $10^{5}$ times for
the given case. 

	 Besides this, the amplification of probe laser radiation 
was observed. Taking into account, that in some of these 
experiments the number of the molecules, exited by the first 
laser, did not exceed $\sim 0,1 \%$, it is a typical case of 
amplification without inversion. Thus the cross-section of 
the reversed optical transition should be at least on three 
orders of magnitude more, than of forward one. Because of the 
Einstein coefficients for forward and reversed transitions should 
be equal, the present estimation for cross-section of the reversed
transition is, obviously, underestimated.

	 Now these experiments are the unique direct and complete 
experimental proof of inequality of forward and reversed optical 
transitions. However, the given physical object, probably, is not
so unique. The rather similar experimental results were received 
earlier in a solid state. In homogeneously broadened absorption 
line of a ruby the dip with width only 37 Hz was observed [26,27]. 
And in this case the huge homogeneous width of optical transition 
is combined with the big lifetime of the exited state toward to 
spontaneous emission ($\sim 3,4 ms $). The measurement of width 
of the dip was carried out by amplitude modulation of laser 
radiation. Unfortunately usual pump-probe experiments with using 
two pulse lasers were not carried out. Such experiments have not 
 so high spectral resolution. But such results can have more 
evident, convincing and direct physical interpretation. 

\section{ Nonlinear optical effects.}

	 The traditional explanation of a nature of nonlinear optical 
processes has descriptive character. Their existence are connected 
with so-called nonlinear susceptibility of atoms and molecules [28]. 
If appropriate factor of a nonlinear susceptibility is great enough,
the process, for example, of a  wave mixing can occur. If this 
factor is equal to zero, the wave mixing is absent. Also, a 
restriction on the lowest order wave mixing process exists, that 
is connected with the symmetry of the process. Namely, the factor 
of nonlinear susceptibility ${\bf\chi^2}$ is nonzero only in a 
medium without center of inversion.

	 Within the framework of discussed representations the main 
reason of nonlinear optical processes is the inequality of forward 
and reversed optical transitions. As the result the atoms and 
molecules have some "memory" about their initial state and 
aspiration to return back. A basis of this aspiration is the high 
cross-section of optical transition into the initial state. And it 
is necessary to understand that the concept of initial state should 
include not only concrete quantum energy level, but also the 
orientation of atom in space toward the laser beam and even a 
phase of vibrational motion of atoms in a molecule.

	 In this case it becomes clear why in gas and liquid phases 
there are no effective processes of three and five photon mixing. 
Every photon has spin. Therefore, using odd number of photons it 
is impossible to return to an initial state. At the same time in 
crystal lattice, where the rotation can be suppressed, such 
processes are possible.

	 But in gas and liquid phases there are a huge amount of four
photon mixing processes. The set of nonlinear optical effects are 
possible and necessary to consider as result of four photon mixing 
process [29].

	 Here we shall discuss in detail only one such effect. It is 
one of the most complex effect. Despite of old history and constant 
interest to it, it is insufficiently well experimentally 
investigated till now. The main problem here is that the existing 
theoretical representations about its nature strongly prevents 
experimenters. There is a photon echo effect. 

	 The atoms and molecules, exited by laser radiation, are 
capable to radiate coherently this energy. When it occurs at once 
after laser pulse, such radiation calls as a nutation effect. The 
complete radiation of the reserved energy does not occur. The 
dephasing processes are one of the reasons. Some of the dephasing 
processes can be reversed in time, then a pulse of a photon echo 
superradiation can appear [2]. 

	 Usual two-pulse photon echo arises when environment is 
influenced by two pulses of laser radiation (P1 and P2) with a 
delay $t_{21}$ between them. Then at time $t_{e2} = t_{21}$ after the 
second laser pulse there is a pulse of photon echo. There is also 
variant of stimulated photon echo, when on environment work 
with three laser pulses (P1, P2, P3). Then the pulse of photon 
echo occurs after a pulse P3 with a delay equal delay between 
pulses P1 and P2 ($t_{e3} = t_{21}$). The time of a delay $t_{32}$ in 
this case can be very large.

	 The Bloch equations well describe dynamics of photon echo, 
but nothing says about a physical nature of this process. The 
concepts of optical, Zeeman coherences and superposition of states 
are used for explanation a photon echo effect [30]. However, 
such approach contradicts some experimental facts. So the study 
of stimulated photon echo in a case, when the delay time $t_{32}$ 
between the second and third pulses considerably exceeds the 
lifetime of the excited state of atoms, clearly specifies 
that the information about the photon echo in this period of 
time contains only in a unique ground state [31]. In this case 
such result is a natural consequence of small lifetime of the 
excited state. But such data can be received also in other way. 
Using an additional laser pulse of other frequency [32], it is 
possible to remove population of the ground or excited states to 
any other far level and, thus, to determine in what state 
(excited or ground) is concentrated the information about 
photon echo in each period of time. Such experiments were not 
carried out yet. 

	 Within the framework of concept about inequality of forward 
and reversed transitions in optics for an explanation of a nature 
of the photon echo it is not necessary to attract representations 
about a superposition of states. In each period of time the 
information about a photon echo is stored at one concrete level 
(ground or excited). Discussed in [2] illustration of a photon 
echo as an example of sports competition of runner actually is, 
obviously, absolutely exact physical analog of a photon echo 
process. The run on a circle corresponds to rotation or precession 
of atoms and molecules. The role of "Maxwell's demon" can and 
should carry out the spin of photon. The absorption or emission 
of photon is accompanied by change of a rotational state of atom
or molecule. 

	 The photon echo can be considered as multiple stage process 
of four photon mixing, which includes stages of dephasing and 
rephasing of rotational motion. The corresponding scheme of energy 
levels is given on Fig.4. Here we shall consider common variant, 
when the longitudinal magnetic field is applied and the magnetic 
levels are not degenerated. Under action of radiation of the first 
laser pulse (P1) the absorption of photon and forward transition 
in the excited state takes place. The dephasing process, which is 
connected with inhomogeneity of rotational motion [33], here begins.
Under action of the second laser pulse (P2) of the same frequency 
and polarization the stimulated emission of photon occurs and the 
atom comes back in the ground state. For a part of atoms, which had 
the appropriate orientation in space, this optical transition will 
be reversed. Its will return to an initial state and their "memory" 
will be "deleted". For other part of atoms such transition will be 
a forward one and the "memory" about the initial state will be kept.
The rotation in the ground state for discussed case is absent and 
the dephasing process stops. Under action of radiation of the third 
laser pulse (P3) of other frequency and helicity a photon absorption 
and excitation of atoms in other excited state takes place, where the 
rotation occurs in the opposite side. The same heterogeneity of 
rotational motion results that the rephasing process begins. At 
time equal $t_{21}$ the rephasing process is finished and due to high 
cross-section of the reversed optical transition into the initial 
state the superradiation of a photon echo appears.

	 According to the given physical explanation the information
about the photon echo after the first pulse is stored in the first 
excited state. After the second laser pulse (for a case of stimulated 
photon echo) it is only in the ground state. At last after the 
third pulse this information is concentrated in the other excited 
state. For usual two-pulse photon echo, when the stage between the 
second and third pulses is absent, the information about the photon 
echo during all period of its formation is stored only in the excited 
states. In this case under action of the second laser pulse two 
consistently optical transitions take place.

	 The scheme of Fig.4 was realized in work [34], where the 
shape of a stimulated photon echo in ytterbium vapor was 
experimentally studied. However later this scheme was 
reconsidered. In work [35] for an explanation of the same 
experimental results other experimental scheme was proposed. In 
this case the second pulse P2 has frequency different from 
frequency of the first pulse P1, and the frequency of the third 
pulse P3, on the contrary, coincides with frequency of the first laser 
pulse. The latter scheme, certainly, better corresponds to the 
concepts of coherency and superposition of states, but in this case 
the four-photon mixing is impossible. Therefore we believe, that 
the experimental scheme described in [34], is correct in contrast 
to the scheme, described in [35]. 

	 These interesting experiments in any case require continuation. 
From the text of [34] it is not clear, why the authors used this 
complicated scheme of photon echo. What will happen, if the 
frequency and helicity of the third laser pulse will be the same, 
as for the first two pulses:

1) the signal  of photon echo will be the same,

2) it will be weak,

3) it will be absent?

The phenomenon of a photon echo is one of many nonlinear optical
effects, where the principle of inequality of optical
transitions allows to understand physical sense of proceeding
processes and to plan ways of their further experimental study.

\section {Conclusion.}

Thus, we have quite sufficient experimental proofs of inequality 
of forward and reversed transitions in optics. The major task 
here for the experimenters is to learn to measure the parameters 
of the reversed optical transitions. Before the theorists there 
is a problem of creation the mathematical model (alternative to 
the Bloch equations) for description the dynamics of optical 
transitions. Such model should include a principle of inequality 
of optical transitions [36]. The possible base of such model may 
be the Dirac equation, which in usual variant assumes time 
invariance violation in electromagnetic interactions [37]. The 
description of the physical phenomena with such model will have 
much more direct and clear physical sense, than the existing 
description.

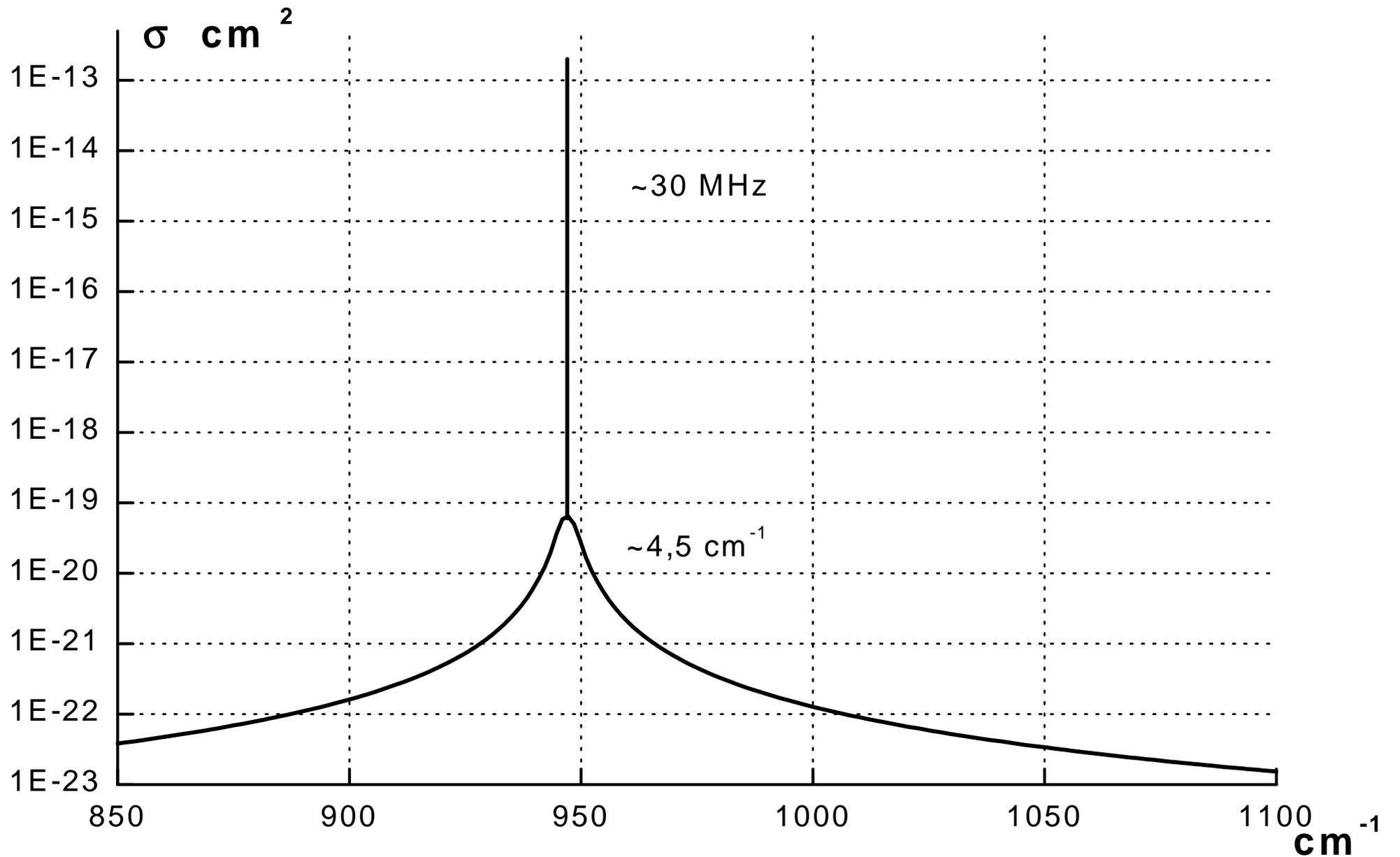
Fig.1 Profile of absorption line of $SF_6$ molecules for the $\nu_3$ band 
$1\leftarrow0$ transition.

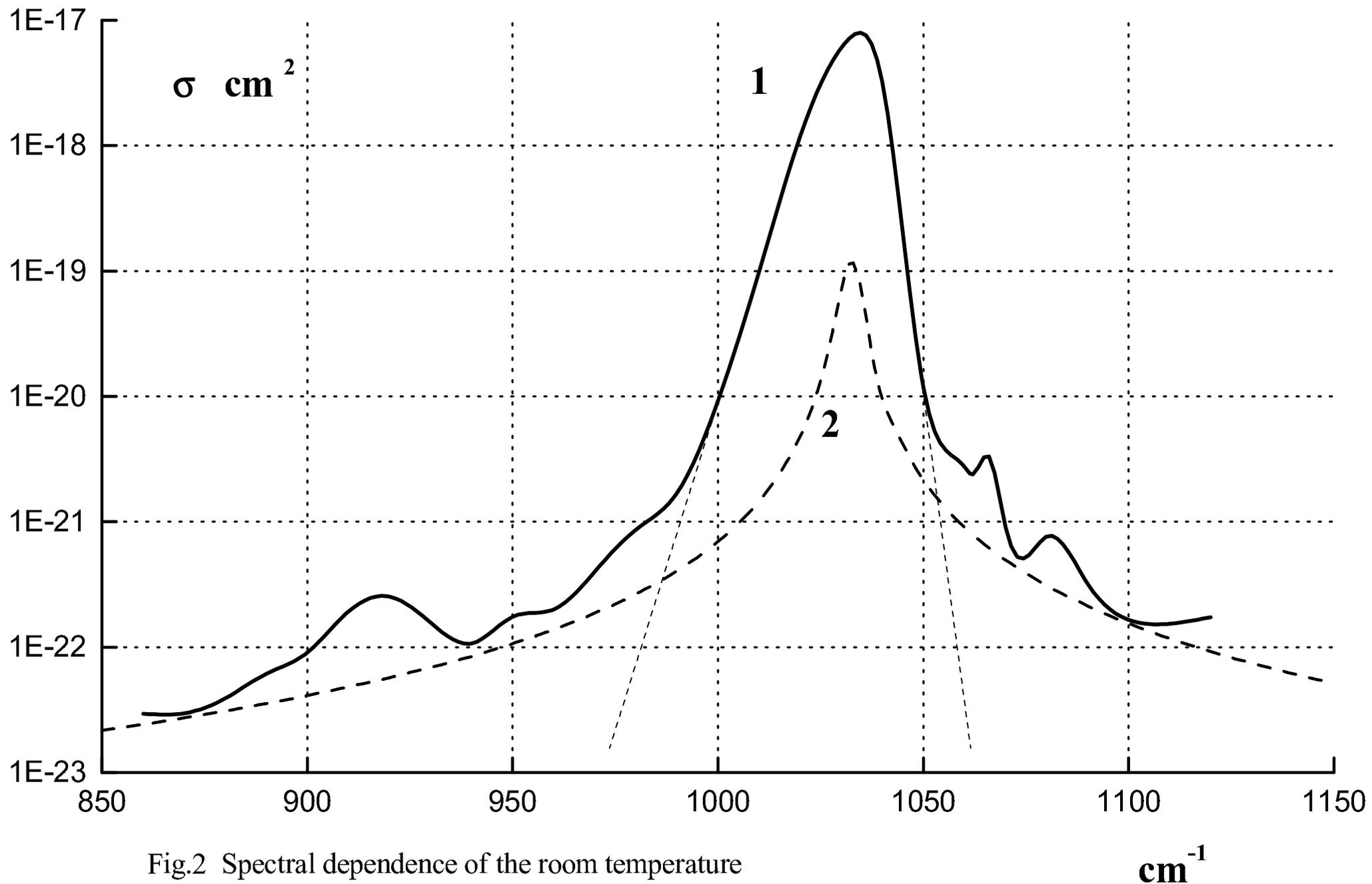
Fig.2 Spectral dependence of the room temperature absorption cross-section
 of $SiF_4$ molecules.
1- spectrophotometer result.
2- Lorentzian profile with FWHM = $4.5\ cm^{-1}$.

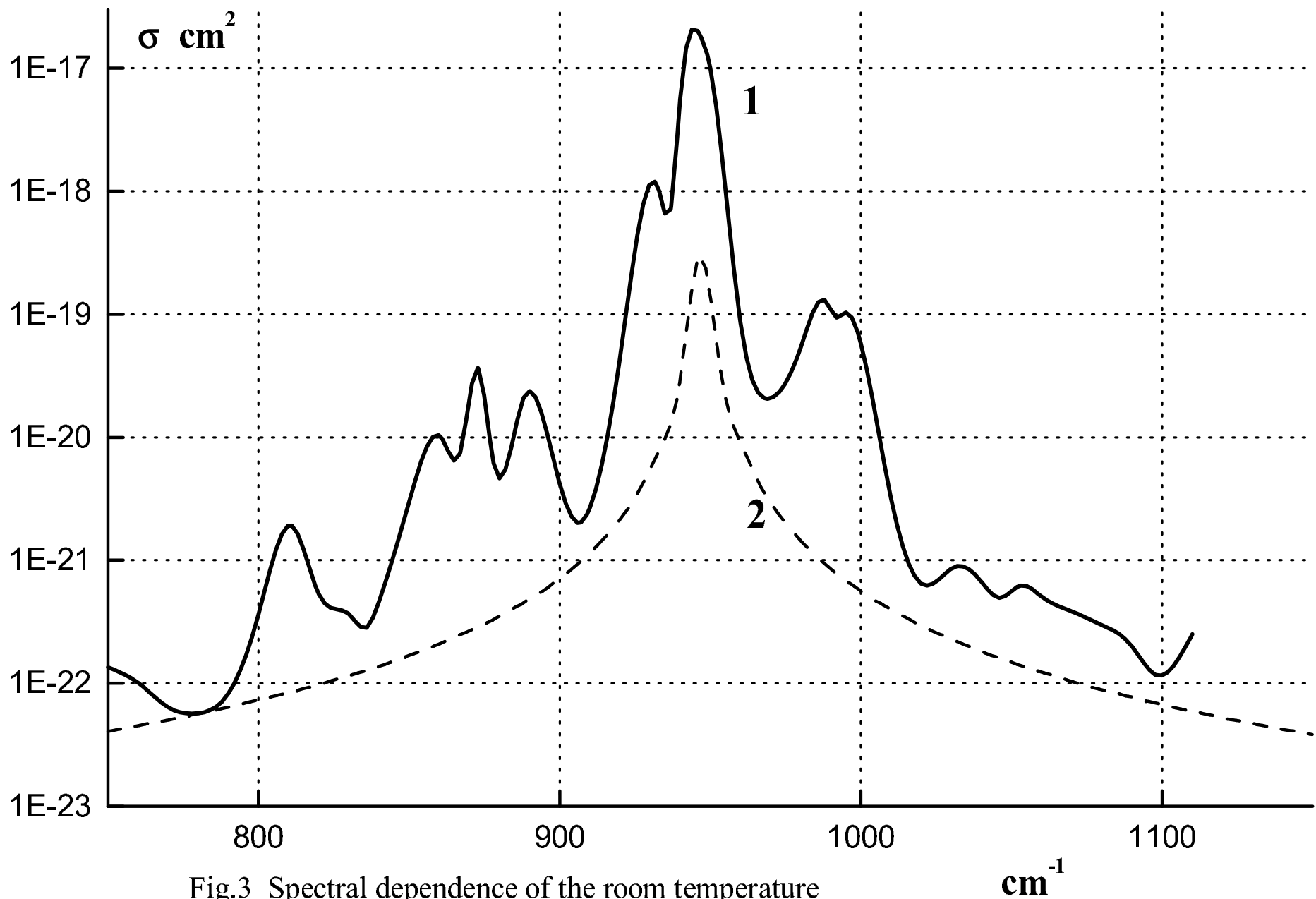
Fig.3  Spectral dependence of the room temperature absorption cross--section
 of $SF_6$ molecules.
1- spectrophotometer result.
2- Lorentzian profile with FWHM = $4.5\ cm^{-1}$.

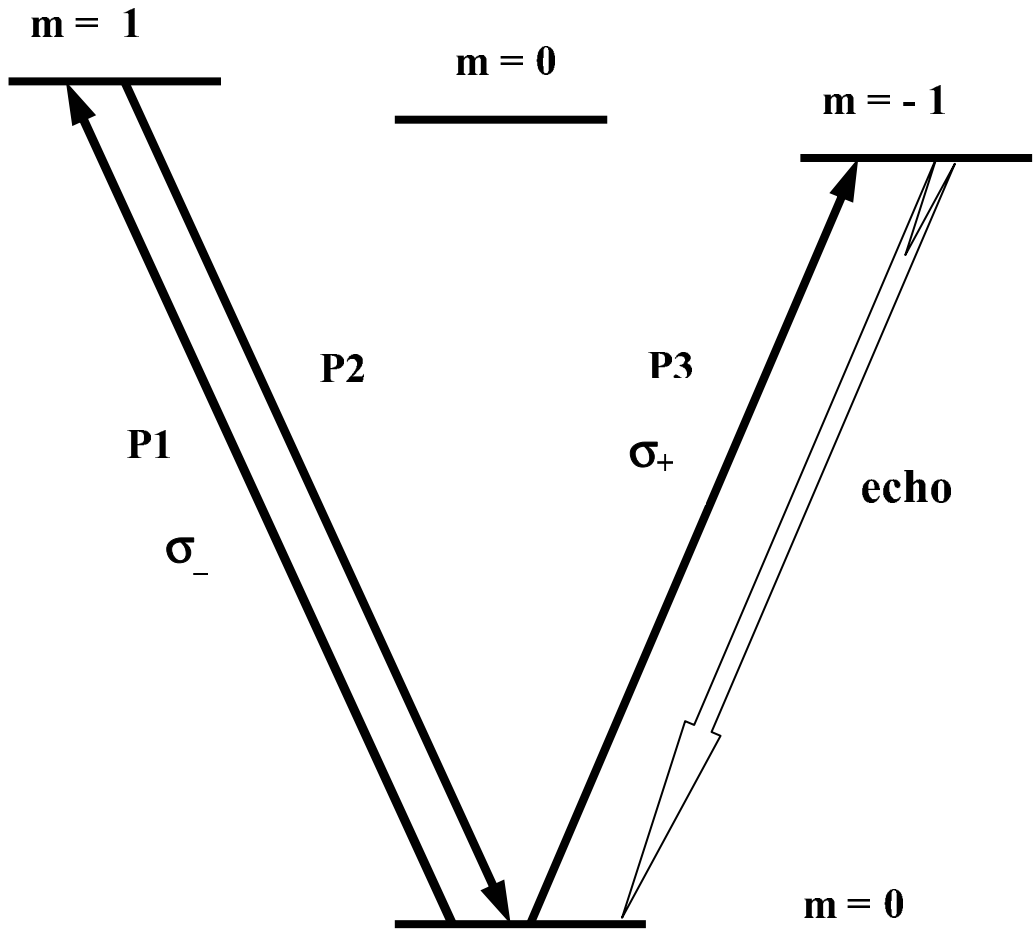
Fig.4   The basic scheme of a photon echo process.

\end{document}